\documentclass[reprint, showpacs, amsmath,amssymb, aps,
prb]{revtex4-1}

\usepackage{mathrsfs,amsmath}
\usepackage{graphicx}
\usepackage{dcolumn}
\usepackage[utf8]{inputenc}%
\usepackage{subfigure}%
\usepackage{bm}
\usepackage{natbib}
\begin{document}


\title{A Surface Admittance Equivalence Principle\\ for Non-Radiating and Cloaking Problems}
%
\author{Giuseppe Labate}
%
\affiliation{
Politecnico di Torino, Department of Electronic and Telecommunications, Torino, 10129, Italy}

\author{Andrea Alù}
\affiliation{
The University of Texas at Austin, Department of Electrical and Computer Engineering, Austin, Texas, 78712, USA}
\author{Ladislau Matekovits}
\affiliation{
Politecnico di Torino, Department of Electronic and Telecommunications,  Torino, 10129, Italy, and with Macquarie University, Sydney, 2109, Australia}%

\date{\today}

\begin{abstract}
In this paper, we address non-radiating and cloaking problems exploiting the surface equivalence principle, by imposing at any arbitrary boundary the control of the admittance discontinuity between the overall object (with or without cloak) and the background. After a rigorous demonstration, we apply this model to a non-radiating problem, appealing for anapole modes and metamolecules modeling, and to a cloaking problem, appealing for non-Foster metasurface design. A straightforward analytical condition is obtained for controlling the scattering of a dielectric object over a surface boundary of interest. Previous quasi-static results are confirmed and a general closed-form solution beyond the subwavelength regime is provided.  In addition, this formulation  can be extended to other wave phenomena once the proper admittance function is defined (thermal, acoustics, elastomechanics, etc.).
\end{abstract}
%
\maketitle
%
%
%
In search of a method for calculating the radiated power from infinitely thin scattering structures, Schelkunoff was led in 1936 to ``certain equivalence theorems'', in order to find the causal relation between arbitrary radiating fields and sources located at a surface boundary \cite{Schelkunoff_1}. In 1938, he also highlighted the concept of impedance for radiating problems as a powerful tool that ``brings out a certain underlying unity in what otherwise appear diverse physical phenomena''  \cite{Imp_Gen}. 
In 1973, Devaney and Wolf established necessary and sufficient conditions for localized sources with special arbitrary fields, mainly \textit{non-radiating} outside their domain of definition, according to a set of theorems they rigorously derived \cite{Dev_Wolf}.
In recent years, such non-radiating sources, difficult to be excited naturally in a bare particle, have been impressed artificially through the insertion of a properly designed cloak \cite{PC,TO1,TO2,MC}, attached or detached to the original scatterer (uncloaked). 

In this framework, exploiting the surface equivalence principle, we impose the design of non-radiating sources in volumetric domains as projected at an arbitrary surface boundary, enclosing the bare particle (non-radiating problem) or the uncloaked object with its coating layer (cloaking problem).  Recently, the surface equivalence principle has been applied to the synthesis of planar devices with reflectionless sheets \cite{Grbic} and to cloak conformal structures with antenna elements \cite{Eleft}. 

Instead of reasoning on the sources as discontinuities of tangential fields at a thin surface \cite{Schelkunoff_1}, we  show that all the useful information relevant to the volumetric interactions between the object (without or with the surrounding cloak) and the background can be recasted in terms of fields ratio over a surface of choice, directly providing admittance functions: this concept is valid at any frequency regime and without any approximation as highlighted by Schelkunoff  \cite{Imp_Gen}. 

For cloaking problems, several approaches exist in the literature for the design of the coating layer according to the size and/or constitutive parameters of the object to be hidden, such as  Plasmonic Cloaking (PC) \cite{PC}, Transformation Optics (TO) \cite{TO1,TO2} and Mantle Cloaking (MC) \cite{MC,Patt_Meta,Fost_Meta}. The TO technique is based on a spatial transformation of fields, preserving their free-propagating characteristic outside a certain region (exact zero scattering), while compressing wave propagation in the annular cloaking medium, through its anisotropic layout, rerouting the energy flow \cite{TO1,TO2}. In such a way, a hole for the fields is created, with the possibility of hiding whatever object inside the cloak once defined its size, regardless the constitutive parameters of the object itself. 
\begin{figure}[h!]
\centering
\includegraphics[width=0.48\textwidth]{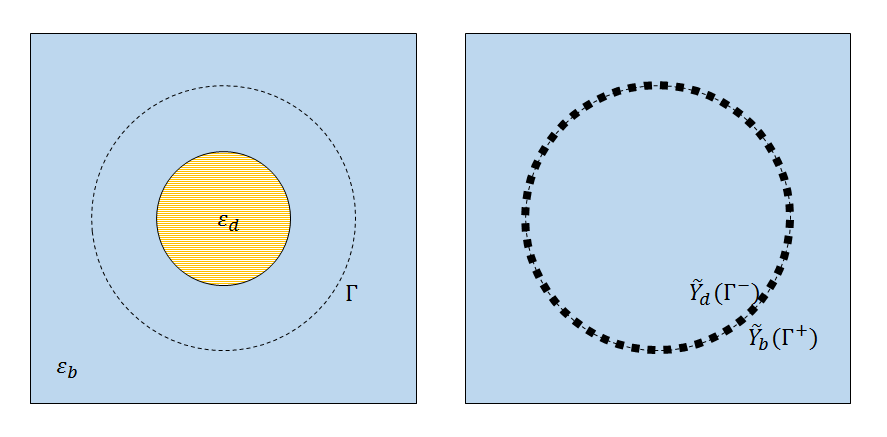}
\caption{\small Non-radiating problem: volumetric scattering from a bare dielectric particle with permittivity $\varepsilon_d$ in a background medium $\varepsilon_b$ (right) and equivalent treatment at an arbitrary boundary $\Gamma$ with normalized admittances $\widetilde{Y}_d(\cdot)$ for the dielectric object at $\Gamma^-$ and $\widetilde{Y}_b$ for the background material at $\Gamma^+$ (right).}
\label{fig:nonrad}
\end{figure}

The PC and MC approaches are based on the scattering cancellation method \cite{PC,MC}, where, by taking into account the scattering from the uncloaked object, the outgoing scattered fields are turned-off to zero or to very low values by a bulk plasmonic or volumetric metamaterial coating (PC)\cite{PC} or, in case of the MC approach \cite{MC}, by  a thin  metasurface, matemathically modeled as an impedance sheet $Z_s$. 

The use of such surface impedance cloaks turns out to be useful if also related to what Schelkunoff reported in his paper on the impedance concept \cite{Imp_Gen}, about ``the idea of extending the $V/I$  relation (voltage-current ratio) from circuits to radiation fields''.

In order to establish a complete non-radiating and cloaking condition valid at any frequency regime, we exploit the surface equivalence principle\cite{Schelkunoff_1} combined with the impedance concept \cite{Imp_Gen}, by considering the bare object (without or with cloak) and the background in terms of their admittance functions relative to a specific incoming wave: this concept is generally exploited in static or quasi-static models (mainly circuit problems), but it is  also implicitly contained in Lorenz-Mie theory \cite{Bohren}. 
In this work, the problem is particularized to electromagnetics, even though such methodology can be extended in a straightforward manner to any other physical scattering phenomenon, once the admittance function is properly defined, such as in elastic  \cite{Elastic}, thermal \cite{Thermal} or acoustic  \cite{Acoustic} non-radiating and cloaking problems. 

As a starting point, we reconsider one of the two classes of non-radiating solutions as derived from the Devaney-Wolf theorem \cite{Laby_opex}. According to theorem III as originally derived \cite{Dev_Wolf}, a necessary and sufficient condition for a physical (or equivalent) surface source distribution $\vec{Q}_s$, projected along its unit polarization vector $\hat{q}$, to be non-radiating is
\begin{align}
\int_{\Gamma} \begin{bmatrix}\vec{Q}_s(\rho) \cdot \hat{q}\end{bmatrix} e^{-\jmath k_b \rho} d\Gamma=0 \hspace{3mm} \mbox{with $\rho \in \Gamma$}
\label{theorem}
\end{align}
where $k_b$ is the wavenumber in the background medium. By ensuring to zero all its Fourier components, theorem \eqref{theorem} predicts the existence of non-radiating sources. An apparently-trivial solution that satisfies the Devaney-Wolf theorem is  
\begin{align}
 Q_s(\rho=\Gamma)=\begin{cases}\vec{J}_s(\Gamma)=0\\
\vec{M}_s(\Gamma)=0\end{cases}
\label{strong}
\end{align}
where the general surface sources $Q_s(\rho=\Gamma)$ have been explicited in terms of $\vec{J}_s$ and $\vec{M}_s$, electric and magnetic surface current densities. Possessing all local zeros in the domain where the source itself is localized (in this case, the surface boundary $\Gamma$), this configuration has been referred as \textit{strong solution} for non-radiating and cloaking problems \cite{Laby_opex}, due to its vanishing components at any considered point. Because of theorem \eqref{theorem} is applied to equivalent rather than physical sources, this kind of solution leads to non-trivial results.

As rediscovered by Schelkunoff \cite{Schelkunoff_1}, a general radiating event, supported by physical scatterers, can be represented and replaced by electric and magnetic surface equivalent sources located at an arbitrary boundary $\Gamma$ as  
\begin{align}
\vec{J}_s(\Gamma)= \hat{n} \times \begin{bmatrix} \vec{H}(\Gamma^+)-\vec{H}(\Gamma^-)\end{bmatrix}\\
\vec{M}_s(\Gamma)= -\hat{n} \times \begin{bmatrix} \vec{E}(\Gamma^+)-\vec{E}(\Gamma^-)\end{bmatrix}
\end{align}
where  $\vec{J}_s$ and $\vec{M}_s$ surface sources, located at $\Gamma$, are proportional to the discontinuity of tangential magnetic and electric fields at the outer ($^+$) and the inner ($^-$) side of the surface boundary $\Gamma$. If such independent electromagnetic sources are identically zero at $\Gamma$, according to Eq. \eqref{strong}, a non-radiating and cloaking condition is thus achieved, giving
\begin{align}
\label{cond1}
\vec{H}(\Gamma^+)=\vec{H}(\Gamma^-)\\
\vec{E}(\Gamma^+)=\vec{E}(\Gamma^-)
\label{cond2}
\end{align}
The simultaneous conditions in Eq. \eqref{cond1}-\eqref{cond2} can be rewritten in a compact form, using the admittance functions as defined in terms of magnetic-electric fields ratio, forming for each component the condition
\begin{align}
\widetilde{Y}_b(\Gamma^+)-\widetilde{Y}_d(\Gamma^-)=0
\end{align}
which aims to control the ratio of the magnetic and electric tangetial fields (normalized admittance functions), through properly incorporating the scatterers at the outer (background material) and the inner side (dielectric material) of the surface boundary $\Gamma$.

Consider for simplicity, as shown in Fig. \ref{fig:nonrad}, a dielectric cylinder of absolute permittivity $\varepsilon_d$, permeability $\mu_b$ and circular transverse section of radius $a$ in an infinite background medium of permittivity $\varepsilon_b$ and permeability $\mu_b$. 
Due to the arbitrary choice of the surface boundary, we choose  $\Gamma$ to be directly attached at $\rho=a$ to the surface of the bare dielectric cylinder, positioned with axis parallel to $\hat{z}$ and illuminated by an incoming TM$_z$ polarization wave (the largest contribution to scattering for dielectrics). In this scenario, the total fields can be analytically computed using  Lorenz-Mie theory \cite{Bohren}. 
Even if the object is still without any cloak, there is a non-radiating condition for which the coefficient $c_n^{\mbox{TM}}$ vanishes or becomes near-zero for a specific scattering order, without any coating. Applying Cramer's rule, such condition reads 
\begin{align}
\begin{vmatrix}
 J_n(k_d a) &\ J_n(k_b a) \\ &\  \\   k_d  J^{'}_n(k_d a) &\ k_b J^{'}_n(k_b a)\end{vmatrix}=0
\end{align}
which is valid for each harmonic index $n$.
Solving this determinant and rewriting such relation in terms of admittance functions, the non-radiating condition reads
\begin{align}
-i\dfrac{J^{'}_n(k_b a)}{J_n(k_b a)} +i\sqrt{\varepsilon_r}\dfrac{J^{'}_n(k_d a)}{J_n(k_d a)} =0 
\label{uncl_eq}
\end{align}
where $\varepsilon_d\equiv\varepsilon_r\varepsilon_b$.
The first term can be recognized as the normalized admittance $\widetilde{Y}_b(\Gamma^+)$ (ratio of normalized magnetic and electric field functions) as computed at $\rho=a^+$ for the specific cylindrical harmonic of choice in a free-space (complete background) scenario \cite{Marcuvitz}. This is consistent with Schelkunoff's idea of considering the impedance/admittance ``as an attribute of the field as well as of the body or the medium which supports the field, so that the impedance to a plane wave is not the same as the impedance to a cylindrical wave, even when both are propagated in infinite free-space'' \cite{Imp_Gen}. The same reasoning can be applied to the second term,  recognized to be the normalized admittance $\widetilde{Y}_d(\Gamma^-)=-i\sqrt{\varepsilon_r} J^{'}_n(k_d a)/J_n(k_d a)$ for each incoming cylindrical harmonic when traveling in a medium with $\varepsilon_r$ as relative permittivity \cite{Marcuvitz}. Beyond the trivial case when $\varepsilon_r=1$, Eq. \eqref{uncl_eq} can be
therefore interpreted as the fact that the scattering due to a certain harmonic can be minimized when the admittance (or impedance) of the two media (bare particle and background) is the same at the surface of the object for the harmonic of interest.

As shown in Fig. \ref{fig:nonrad}, this interpretation of a homogeneous dielectric particle in a volumetric non-radiating condition, which can show weak scattering responses even without any loading surface for some frequency values, can serve as surface admittance model for dielectric nanoparticles supporting a non-radiating  \textit{anapole} mode and \textit{metamolecules} \cite{ana1,ana2}.

Once reconsidered the non-radiating volumetric problem at the surface boundary $\Gamma$, it becomes straightforward to solve also the cloaking problem.

Eq. \eqref{uncl_eq} can now be altered and controlled through the insertion of a proper cloaking surface impedance, or mantle cloak, which is an additional normalized admittance sheet $\widetilde{Y}_s$: the intentional choice of $\Gamma$, to be directly attached to the surface of the dielectric cylinder, leads the fictitious equivalent zero surface sources to be sustained and implemented by a physical dispersive cloak. Due to the insertion of such lumped surface admittance in parallel, the relation in Eq. \eqref{uncl_eq} is modified into
\begin{equation}
\widetilde{Y}_b(k_b a, n)-\begin{bmatrix}
\widetilde{Y}_d(k_b a,\varepsilon_r, n)+\widetilde{Y}_s\end{bmatrix}=0
\label{zero_contrast}
\end{equation} 
where the dependence from the three
main dimensionless variables of the cloaking problem are: the normalized size with respect to the background wavelength $k_b a=2\pi a/\lambda$, the relative permittivity of the uncloaked object $\varepsilon_r$ and the cylindrical harmonic index $n$.  
Similar to Eq. \eqref{uncl_eq}, we can expect that the proper tailoring of $\widetilde{Y}_s$ in Eq. \eqref{zero_contrast} is able to match the two impedances on the surface of the object, and therefore suppress the scattering contribution from harmonic $n$.

\begin{figure}[ht!]
\centering
\includegraphics[width=0.3\textwidth]{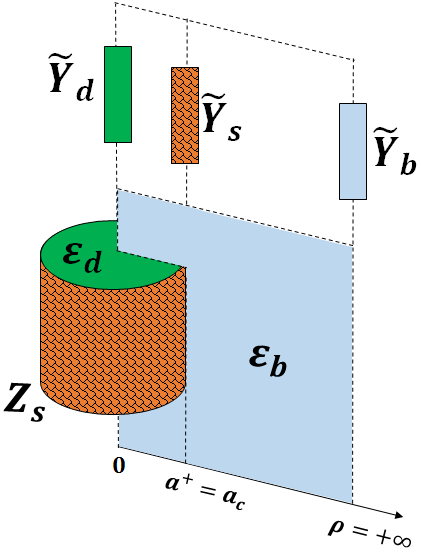}
\caption{\small Cloaking Problem: cloaked dielectric cylinder, having permittivity $\varepsilon_d$ and radius $a$, with surface impedance $Z_s$, in a background region $\varepsilon_b$ (bottom). Formulation in terms of normalized admittances $\widetilde{Y}_d$ (dielectric), $\widetilde{Y}_s$ (cloak) and $\widetilde{Y}_b$ (background), all interacting as lumped elements (top).}
\label{fig:cases}
\end{figure}

Without solving the entire Lorenz-Mie scattering for the overall cloaking problem \cite{MC,Patt_Meta} but perfectly consistent with, the analytical formula \eqref{zero_contrast} predicts the required cloaking impedance $Z_s=Z_{B}/\widetilde{Y}_s=R_s+iX_s$, where $Z_{B}=\sqrt{\mu_b/\varepsilon_b}$ is assumed as the reference background impedance with respect all the other values are normalized to (tilde sign on top), as shown in Fig. \ref{fig:cases}. 

In essence, Eq. \eqref{zero_contrast} exploits the equivalence principle as applied to non-radiating problems: if the scattered field on a surface is identically zero, being  sustained by the absence of sources, it should be zero even everywhere outside the surface boundary $\Gamma$. Therefore, by making sure that the impedance is matched on the surface of interest, which is achieved adding a suitably designed mantle cloak satisfying Eq. \eqref{zero_contrast}, we can make sure that, for the harmonic of interest, the scattering is zero everywhere.

Once defined the constitutive parameter $\varepsilon_r$ and the radius $a$ of the object to be hidden, the residual scattering for each harmonic can be associated to the imaginary part  of the difference between the background and the dielectric object admittances. Considering the frequency regime $\lambdabar\equiv k_b a$ and each $n$ cylindrical harmonic, 
%
%
we define the function $\Delta(\lambdabar,n)$ as 
\begin{equation}
-i\Delta(\lambdabar,n)\equiv\widetilde{Y}_b(\lambdabar, n)-\widetilde{Y}_d(\lambdabar,\varepsilon_r, n) 
\end{equation} 
When such residual function is different from zero for the bare particle, a surface load is required in order to achieve a cloaking effect for the harmonic of interest and it is found to be equal to $\widetilde{Y}_s(\lambdabar,n)=-i\Delta(\lambdabar,n)$ or explicitly 
\begin{equation}
\widetilde{Y}_s(\lambdabar,n)=-i\begin{bmatrix}
\dfrac{J^{'}_n(\lambdabar)}{J_n(\lambdabar)} -\sqrt{\varepsilon_r}\dfrac{J^{'}_n(\lambdabar\sqrt{\varepsilon_r})}{J_n(\lambdabar\sqrt{\varepsilon_r})} \end{bmatrix}\hspace{1mm} \forall n, \forall \lambdabar
\label{zero_contrast_Ys}
\end{equation} 
which is exactly the value needed to compensate such residual difference or \textit{mismatch}.

Interestingly, it is only over a single surface that each $\widetilde{Y}$ function (bare object, cloak and background) takes into account both information about the field shape and material constitutive parameters of the object: consistently with the surface equivalence principle  \cite{Schelkunoff_1}, this ensures the cloaking effect to be achieved at any arbitrary distance far away from the initial surface boundary. 
%
%

Eq. \eqref{zero_contrast_Ys} is consistent with the closed-form analytical formula previously derived in the subwavelength (or quasi-static) frequency regime for mantle cloaks \cite{MC,Patt_Meta,Fost_Meta}. In the conformal case for $\gamma\equiv a/a_c=1$, the value of the normalized surface admittance $\widetilde{Y}_s^{QS}=\widetilde{G}_s^{QS}+i\widetilde{B}_s^{QS}$ in quasi-static regime can be obtained directly from Lorenz-Mie theory as\cite{MC,Patt_Meta}
\begin{align}
X_s^{QS}=+\dfrac{2}{\omega a \varepsilon_b (\varepsilon_r-1)} \hspace{2mm}  \mbox{or} \hspace{2mm} \widetilde{B}_s^{QS} =- \lambdabar\dfrac{(\varepsilon_r-1)}{2} 
\label{qs_formula} 
\end{align}
%
 
The same result is confirmed as a particular case in Eq. \eqref{zero_contrast_Ys}, solved for the dominant mode $n=0$ in the quasi-static frequency regime: in the small argument limit for $x \rightarrow 0$, the Bessel functions become $J_0(x)\approx 1$ and $J^{'}_0(x)\approx -x/2$, thus the result reads
\begin{align}
\widetilde{Y}_s (\lambdabar \ll 1, n=0)=-i\begin{bmatrix}
-\dfrac{\lambdabar}{2}+\sqrt{\varepsilon_r}\dfrac{\lambdabar \sqrt{\varepsilon_r}}{2}\end{bmatrix}=i\widetilde{B}_s^{QS}
\end{align}
%
%
\begin{figure}[ht!]
\centering
\subfigure[]
{\includegraphics[width=0.4\textwidth]{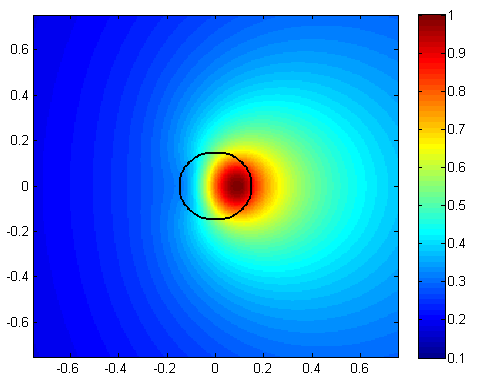}}\\
\subfigure[]
{\includegraphics[width=0.4\textwidth]{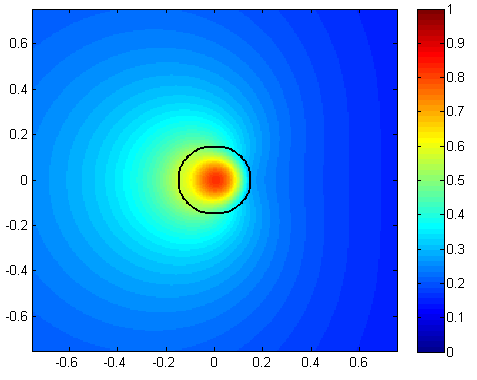}}\\
\subfigure[]
{\includegraphics[width=0.4\textwidth]{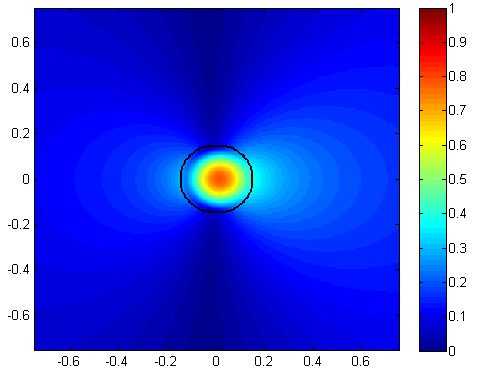}}
\caption{\small Absolute value of the scattered field (with axis normalized to $\lambda$) for the frequency regime value $\lambdabar=0.3\pi\approx 0.94$:  (a) uncloaked dielectric object, (b) cloaked device with $Z_s^{QS}=+i 400\ \Omega$ and  (c) cloaked system with $Z_s^{opt}=+i 216.80\ \Omega$.}
\label{fig:escat}
\end{figure}
We consider now a couple of examples of cloaking for cylinders, in order to highlight the efficient analytical design of mantle cloaks based on this formula. Consider first a dielectric cylinder with $a=0.15\lambda$ (thus, $\lambdabar=0.3 \pi$),  possessing a relative permittivity $\varepsilon_r=3$ with respect to the background $\varepsilon_b=\varepsilon_0$ (free-space, thus $Z_B=Z_0=120\pi\ \Omega$), loaded by a surface impedance at $a_c=a$ to achieve an optimal scattering suppression. Using classical Lorenz-Mie Theory as applied to the uncloaked cylinder plus the impedance sheet, the analytical result in quasi-static condition, using Eq.  \eqref{qs_formula}, leads to the normalized value $\widetilde{Y}_s^{QS}=-i 0.3\pi$ ($Z_s^{QS}=+i400\ \Omega$). We can now analytically derive the optimal value of required surface impedance by using Eq. \eqref{zero_contrast_Ys}. Information about the dominant mode for a certain frequency regime $\lambdabar$ can be derived in a straightforward manner in terms of $\Delta(\lambdabar,n)$, which represents a dimensionless quantity that remains real for lossless scatterers and backgrounds: it is expected that, the largest the mismatch, the largest the contribution (dominant) into the outgoing scattered field for the $n$ cylindrical harmonic mode. For this reason, the strategy here adopted to build the complete dispersion response of the normalized surface admittance is
\begin{align}
\widetilde{Y}_s^{opt}(\lambdabar)=-i\ \underset{n}{\mbox{max}} \ \Delta(\lambdabar,n) 
\label{build_disp}
\end{align}
For monochromatic illumination, the value of the normalized admittance gives a surface impedance value of  $Z_s^{opt}=+i216.80\ \Omega$ in the frequency regime $\lambdabar=0.3 \pi$, for which the first dominant mode to be suppressed appears to be $n=0$.
In Fig. \ref{fig:escat}, the three cases are shown in terms of absolute value of scattered fields (here, the incoming field is completely polarized parallel to the cylinder's axis): the uncloaked dielectric object (a), the cloaked device with $Z_s=Z_s^{QS}$ (b) and the cloaked system with $Z_s=Z_s^{opt}$ (c). 
\begin{figure}[ht!]
\centering
\includegraphics[width=0.4\textwidth]{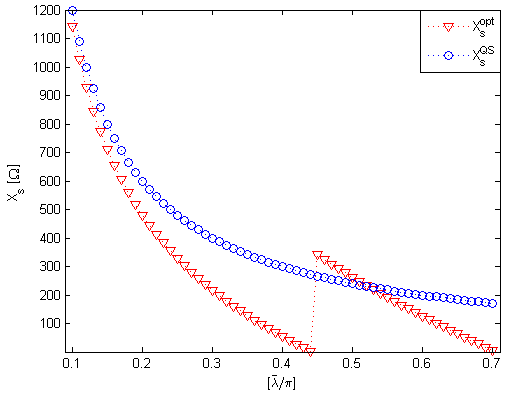}
\caption{\small Dispersion behaviour of $X_s$, imaginary part of $Z_s=iX_s$.  Comparison between quasi-static approximation (QS) and dynamic analytical formulas.}
\label{fig:disp}
\end{figure}

For a monochromatic incoming field, with TM$_z$ polarization, traveling from the left to the right of each panel, the scattered field is maximum for the uncloaked dielectric whereas it is clearly reduced for the two cloaked devices: for the impedance sheet with $Z_s^{QS}$, reflections exist in the backward direction, whereas for the impedance coating with $Z_s^{opt}$ very low scattered energy in the outside region is observed.

For wideband incoming signals, the optimal $Z_s$, once fixed the geometrical and constituive parameters of the cloaked system, can be analytically derived  from the  complete function $\widetilde{Y}_s(\lambdabar)$ using Eq. \eqref{build_disp}. In order to suppress the first dominant mode at any frequency, it is expected that, as  frequency changes, also the first dominant mode moves its index $n$, with consequent jumps in the dispersion of $X_s$ as obtained in Eq. \eqref{build_disp}. 
\begin{figure}[ht!]
\centering
\includegraphics[width=0.4\textwidth]{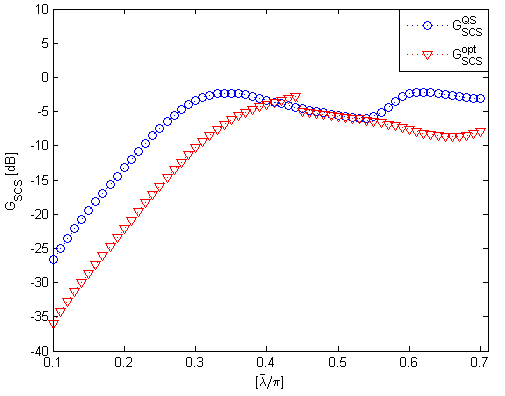}
\caption{\small Gain functions (dB units) as defined in Eq. \eqref{eq:g_scs}. Quasi-static approximation formula (blue point curve) and optimal analytical formula (red point curve), with the dispersion of Fig. \ref{fig:disp}. As a reference, the uncloaked case is the $0$ dB line.}
\label{fig:SCS}
\end{figure}

In order to validate this effect, the dispersion of the surface impedance $Z_s^{opt}(\lambdabar)$ is depicted as a function of the normalized diameter $D_\lambda=\lambdabar/\pi$ in the range $[0.1\div 0.7]\lambda$ for the same cylindrical object. Interestingly, such functional dependence of the surface impedance with respect to the inverse of $\lambdabar$ (thus, directly proportional to $\omega$) is monotonically decreasing: in order to realize the dispersion of such admittance/impedance cloak, which would realize a broadband cloaking device, the Foster's reactance theorem has to be broken \cite{Foster} and non-Foster metasurfaces, loaded with active elements \cite{Fost_Meta}, have to be employed. For this reason, this formulation is appealing to explore the limitations as dictated by all passive Foster cloaks \cite{Mont_inv_exp}.

As shown in Fig. \ref{fig:disp}, a jump arises when $\lambdabar=0.45\pi$, for which $\Delta(\lambdabar,n=1) >\Delta(\lambdabar,n=0)$ and the first dominant term passes from $n=0$ (the same as in the quasi-static regime) to $n=1$ (and it is mantained up to the final value $\lambdabar=0.7\pi$). The gain in term of SCS for the cloaked scenario is defined, with respect to the uncloaked scenario, as
\begin{align}
G_{SCS}(\lambdabar)=\dfrac{\sum_{n=0}^{N_{max}} \begin{pmatrix} 2- \delta_{n,0}\end{pmatrix} \begin{vmatrix}c_n^{\mbox{TM}}(\lambdabar)\end{vmatrix}^2_{clk} }{ \sum_{n=0}^{N_{max}} \begin{pmatrix}2- \delta_{n,0}\end{pmatrix} \begin{vmatrix}c_n^{\mbox{TM}}(\lambdabar)\end{vmatrix}^2_{unc} }
\label{eq:g_scs}
\end{align}
where $\delta_{n,0}$ is the Kronecker delta, which takes into account the symmetric contribution for $\pm n$ harmonics with respect to the central index $n=0$, as mentioned above.
The SCS gain leads to less-than-unity values (thus a negative sign in dB units) for the $G_{SCS}(\lambdabar)$ function if the scattering of the uncloaked structured is very large with respect to the cloaked case.
As reported in Fig. \ref{fig:SCS}, this is the case for the entire frequency regime window. As also shown in Fig. \ref{fig:escat} at $\lambdabar=0.3\pi$, the improvement of the optimal surface impedance with proper dispersion (red triangle line) is around $-10$ dB with respect to the cloaked case with quasi-static approximation formula (blue point line). Around the frequency regime value $\lambdabar=0.45 \pi \approx 1.41$, the function $G_{SCS}^{opt}(\lambdabar)$ becomes slightly worse with respect to $G_{SCS}^{QS}(\lambdabar)$. This can be due to a good trade-off achieved by Eq. \eqref{qs_formula} not for the cancellation of the entire harmonic $n=0$ but for a minimization of the overall mismatch for $n=0$ and $n=1$ simultaneously. 
\begin{figure}[ht!]
\vspace{5mm}
\centering
\subfigure[]
{\includegraphics[width=0.4\textwidth]{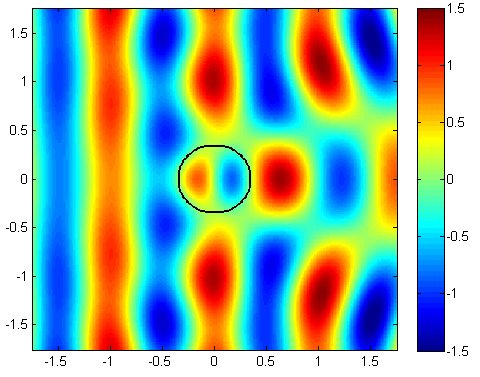}}\\
\subfigure[]
{\includegraphics[width=0.4\textwidth]{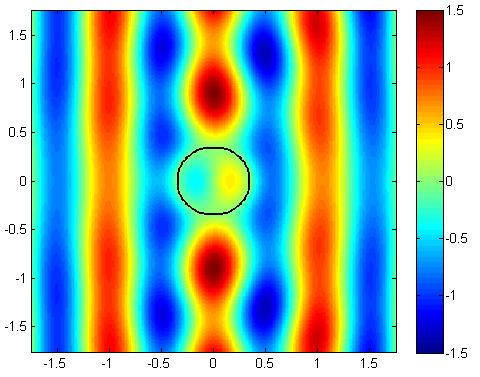}}\\
\caption{\small Real part of the total field (with axis normalized to $\lambda$) for the frequency regime value $\lambdabar=0.7\pi\approx 2.20$:  (left) uncloaked dielectric object and (right) cloaked system with $Z_s^{opt}=+4.93 \Omega$.}
\label{fig:etot}
\end{figure}
From $\lambdabar=0.45\pi$ to $\lambdabar=0.55\pi$, both gain functions are similar, because by chance this corresponds to similar values in the $Z_s$ as reported in Fig. \ref{fig:disp} and, towards the end, while the quasi-static approximation design gets closer and closer to $0$ dB (the uncloaked case), the analytical formula exploited for the first dominant term is able to achieve a drastic reduction around $-6$ dB from $\lambdabar=0.65\pi$ until the end of this frequency regime window. 

The real part of the total electric field is shown for the frequency regime $\lambdabar=0.7\pi  \approx 2.20$ in Fig. \ref{fig:etot}, for the uncloaked and optimal cloaked cases with the first dominant mode ($n=1$) suppressed with $Z_s^{opt}=+4.93\ \Omega$.   
When the overall object size exceeds $\lambdabar \geq 2$, it becomes more and more challenging to reduce the SCS using a single impedance sheet as observed beyond this frequency window (not reported here). This implies that for $\lambdabar \geq 2$ the scattering is dominated by two (or more) different dominant cylindrical harmonics but, due to the fact that each cloaking shell (even in the volumetric case) can control one single harmonic at time, there is at least the direct control for cancellation of only the first dominant term: this does not ensure a reduction in the overall SCS if the second dominant term (or the third, and so on) is comparable with respect to the first contribution. However, a systematic generalization, based on such surface admittance equivalence principle, towards multilayer impedance cloaks is under investigation.  

In conclusion, a reformulation of the surface equivalence principle, in terms of discontinuity in tangential fields components \cite{Schelkunoff_1}, has been proposed in terms of field ratio at the same surface boundary. For non-radiating and cloaking problems,  the overall scattering interactions can be written in terms of the admittance functions for the bare object (without or with cloak) and background, calculated over a surface of choice. For non-radiating problems, the zero surface sources solution of the Devaney-Wolf theorem ensures an admittance model, appealing for anapole modes and metamolecules with weak radiation properties\cite{ana1,ana2}. For cloaking problems, the closed-form solution for the required surface impedance is adjusted to ensure zero scattering for a specific cylindrical harmonic excitation: previous findings, based on Lorenz-Mie scattering theory are confirmed in quasi-static regime. Comparisons, with a deep analysis on the role of the frequency regime $\lambdabar$ in terms of harmonic scattering control, have been also performed, validating this cloaking admittance model, appealing for the design of non-Foster metasurfaces. These findings can be generalized  in a straightforward manner towards multilayer structures and, in addition, for any wave phenomenon once the admittance/impedance concept, as envisioned by Schelkunoff \cite{Imp_Gen}, can be properly defined and applied. 
%
%
%
%

\end{document}